\documentclass[12pt]{article}
\usepackage{amssymb,cite}
\usepackage{epsf}
\usepackage{graphicx}
\usepackage{psfrag}
\usepackage[dvips]{epsfig}

\def\starup#1{\mbox{$\raise1.8ex\hbox{$*$} \kern-.7em#1$}}
\def\krup#1{\mbox{$\raise1.8ex\hbox{$+$} \kern-.7em#1$}}
\def\linup#1{\mbox{$\raise1.9ex\hbox{---} \kern-1.0em#1$}}

\begin{document}

\title{Rare $t$-quark decays $t \to c \; l_j^{+}  \; l_k^{-}$ \, ,
$t \to c \; \tilde{\nu_j} \; \nu_k$ \, in the minimal four color
symmetry model }
\author{P.Yu.~Popov\thanks{E-mail: popov\_p@univ.uniyar.ac.ru} , \,
 A.D.~Smirnov\thanks{E-mail: asmirnov@univ.uniyar.ac.ru}\\
{\small\it Division of Theoretical Physics, Department of Physics,}\\
{\small\it Yaroslavl State University, Sovietskaya 14,}\\
{\small\it 150000 Yaroslavl, Russia.}}
\date{}
\maketitle

\vspace{-5mm}
\begin{abstract}
The rare $t$-quark decays $t \to c \; l_j^{+}  \; l_k^{-}$ ,
 $t \to c \; \tilde{\nu_j} \; \nu_k$ via the scalar leptoquark doublets
are investigated in the minimal four color symmetry model with the
Higgs mechanism of the quark-lepton mass splitting. The partial
widths of these decays are calculated and the total width of the
charged lepton mode $ \Gamma(t \to c \, {l^{+}}' \,
l^{-})=\sum_{j,k} \Gamma(t \to c \, l_{j}^{+}\, l_{k}^{-})$
 and the total width of the neutrino mode
 $ \Gamma(t \to c \, \tilde{\nu}' \, \nu)=\sum_{j,k} \Gamma(t \to c \, \tilde{\nu_{j}} \, \nu_{k})$
are found. The corresponding branching ratios are shown to be
\begin{eqnarray}
&&Br(t \to c \, {l^{+}}' \, l^{-}) \approx (3.5 - 0.4) \cdot 10^{-5}, \nonumber
\\
&&Br(t \to c \, {\tilde{\nu}}' \, \nu) \approx (7.1 - 0.8) \cdot 10^{-5}
\nonumber
\end{eqnarray}
for the scalar leptoquark masses $m_{S}=180 - 250$~GeV and for the
appropriate values $(\sin{\beta} \approx 0.2)$ of the mixing angle
of the model. The search for such decays at LHC may  be of
interest.
\end{abstract}

\newpage

The search for the possible sings of a new physics beyond the
Standard Model (SM) will be one of the goals of the experiments at
LHC and at the other future colliders. Putting LHC into operation
will essentially enlarge the possibilities for the experiments
with
 t-quark, including the search for the non-SM effects in the top physics
 \cite{F, B}. In particular, the LHC with its very
 large top samples (~about $10^{7}-10^{8}$ top quark pairs per year \cite{B}~)
 will allow the experimental investigation of the rare t-quark
 decays which are forbidden or have unobservable small widths in the SM
but can be essentially enhanced in some extensions of the SM. The
detection of such decays an LHC would be an evident signal of new
physics beyond the SM.

 The most investigated rare decays of t-quark are the
FCNC decays $t \to c X, \; X= \gamma, Z, g, H$. These decays are
very suppressed in the SM ($Br_{SM}(t \to c X) \sim 10^{-13}$ and
$\sim 10^{-11}$ for $X= \gamma, Z, H$ and for $X=g$ \cite{EHS, HTTh,
MPS}) but they can be essentialy enhanced
in some extensions of the SM. For example, in the minimal supersymmetric
standard model (MSSM) the branching ratios of these decays can amount to
the values
 $Br_{MSSM}(t \to c X) \sim
10^{-8}, \sim 10^{-6}, \sim 10^{-4}$ for $X= \gamma, Z, \, X=g, \,
X=h^{0}$   \cite{LOY, YL, CHK, DPS, LNR, CFK, GS},
in the two Higgs doublet model they can be enhanced up
to $Br_{2HDM}(t \to c X) \sim 10^{-7}, \sim 10^{-8},\sim
10^{-5},\sim 10^{-4}$ for $X= \gamma, Z, g, h$
  \cite{EHS, ARS, BGS}, some ehancement of these
decays can also take place in the model with additional quark
singlets \cite{AN}.

One of the possible variant of new physics beyond the SM can  be the variant induced
by the four color symmetry between quarks and leptons of Pati-Salam type \cite{PS}.
The immediate consequence of this symmetry is the prediction 
of the gauge leptoquarks which, however, occur to be relatively heavy. 
For example, the most stringent lower mass limit for the vector leptoquarks 
(resulted from the unobservation of the $K^0_L \rightarrow \mu^{\pm} e^{\mp}$ 
decays) is of order of $10^3$ TeV. Such heavy leptoquarks can affect 
the physics at energies of order of or below  $1$ TeV very weakly. 

It should be noted however that in addition to the gauge leptoquarks 
the four color symmetry can predict also the new particles in 
the scalar sector. Thus, in the case of the Higgs mechanism 
of splitting the masses of quarks and leptons the four color symmetry 
in its minimal realization on the gauge group 
$G=SU_{V}(4)\times SU_{L}(2)\times U_{R}(1)$
(MQLS-model \cite{AD1,AD2}) predicts the existence of the two scalar 
leptoquark doublets $S^{(\pm)}$ belonging to the
(15,2,1) - multiplet of the group G. 
These scalar leptoquark doublets together with the other components 
of the (15, 2, 1) - multiplet and with the (1, 2, 1) - doublet are necessary 
\cite{PovSm1}
for  splitting the the masses of quarks from those of leptons by the Higgs 
mechanism and for generating the quark - lepton mass splittings including 
the so large mass splittings as the $b - \tau $ and $t - \nu_{\tau} $ ones.    
Unlike the vector leptoquarks
the scalar leptoquark doublets $S^{(\pm)}$ can be relatively light, with
masses of order of 400 GeV or less, without any contradictions with the 
$K^0_L \rightarrow \mu^{\pm} e^{\mp}$ data or with the radiative correction 
limits   \cite{AD3,PovSm2}.
 Because of their Higgs origin the coupling constants of these
scalar leptoquark doublets with the fermions are proportional to
the ratios $m_{f}/ \eta $ of the fermion masses $m_f$ to the SM
VEV $\eta$. The effects of these scalar leptoquarks in the
processes with the ordinary u-, d-, s- quarks are small because of the
smallness of the corresponding coupling constants ($ m_u/ \eta
\sim m_d/ \eta \sim 10^{-5},  m_s/ \eta \sim 10^{-3}$), whereas
these effects can be significant in c-, b- and, especially,
in top-physics ( $ m_c/ \eta \sim m_b/
\eta \sim 10^{-2} , \, m_t/ \eta \sim 0.7$).

Ones of the possible new effects which can be induced by 
the scalar leptoquark doublets $S^{(\pm)}$ are 
the specific decays of $t$-quark
\begin{eqnarray}
&&t \to c \; l_j^{+} \; l_k^{-} \; , \label{eq:decayl} \\
&&t \to c \; \tilde{\nu_j} \; \nu_k \,  \label{eq:decaynu}
\end{eqnarray}
with the production of $c$-quark with the pairs $l_j^{+} l_k^{-}$
$j,k=1,2,3$ of charged leptons $l_{k}^{-}= e^{-}, \, \mu^{-}, \,
\tau^{-}$ and antileptons $l_{j}^{+}= e^{+}, \, \mu^{+}, \,
\tau^{+}$, in general of the different generation, or with the
neutrino-antineutrino pairs $\tilde{\nu_j} \nu_k$, $\nu_k$
 are the mass eigenstates of
neutrinos, $\tilde{\nu_j}$ are the antineutrinos. The decays
(\ref{eq:decayl}) in general are different from the generation
diagonal  decays $t \to c X \to c \, l^{+}_j \, l^{-}_j$ predicted
in models mentioned above \cite{EHS, HTTh, MPS, LOY, YL, CHK, DPS,
LNR, CFK, GS, ARS, BGS, AN} and the detection of the decays
(\ref{eq:decayl}), (\ref{eq:decaynu}) would be the signal of the
new physics, possibly, induced by the four color symmetry between
quarks and leptons.

In the present Letter we calculate the contributions of the scalar
leptoquark doublets into decays  (\ref{eq:decayl}),
(\ref{eq:decaynu}) in frame of the minimal four color quark-lepton
symmetry model with the Higgs mechanism of splitting the masses of
quarks and leptons(MQLS-model \cite{AD1, AD2}) and evaluate and
discuss the widths and the branching ratios of these decays in
dependence on the scalar leptoquark masses and on the mixing
parameters of the model.

In MQLS model the basic left ($L$) and right ($R$) quarks
${Q'}^{L,R}_{ia\alpha}$ and leptons ${l'}^{L,R}_{ia}$
form the fundamental quartets of $SU_{V}(4)$ color group and can
be written, in general, as superpositions
\begin{eqnarray}
{Q'}^{L,R}_{ia\alpha}=\sum_{j}(A^{L,R}_{Q_a})_{ij} \, Q^{L,R}_{ja\alpha} ,
\,\,\,\,\,\,\,
{l'}^{L,R}_{ia}=\sum_{j}(A^{L,R}_{l_a})_{ij} \, l^{L,R}_{ja} \; \label{eq:fmix}
\end{eqnarray}
of the quark and lepton mass eigenstates $Q^{L,R}_{ia\alpha}$ ,
$l^{L,R}_{ia}$, where $i,j=1, \, 2, \, 3$ are the generation indexes,
$ a = 1, 2 $ and $ \alpha = 1, 2, 3 $ are the $ SU_{L}(2) $  and
$ SU_{c}(3) $ indexes,
$Q_{i1} \equiv u_i=(u,c,t)$, $Q_{i2} \equiv d_i=(d,s,b)$ are
the up and down quarks, $l_{j1} \equiv \nu_{j}$
are the mass eigenstates of neutrinos and
$l_{j2} \equiv l_{j}=(e^{-}, \mu^{-}, \tau^{-})$
are the charged leptons.
The unitary matrixes $A^{L,R}_{Q_a}$ and $A^{L,R}_{l_a}$ describe
the fermion mixing and diagonalize the mass matrixes of quarks and
leptons.

The scalar leptoquark doublets $S^{(\pm)}$ have the SM hypercharge
$ Y^{SM}_{\pm} = 1 \pm 4/3 $ and can be written in the form
\begin{eqnarray}
S^{(\pm)}_{a \alpha} =
\left ( \begin{array}{c}
S_{1 \alpha}^{(\pm)}\\
S_{2 \alpha}^{(\pm)}
\end{array} \right )  ,
\label{eq:SpSm}
\end{eqnarray}
where the up ($a=1$) leptoquarks $S_{1 \alpha}^{(\pm)}$ have electric
charge $5/3$ and $1/3$ and the down ($a=2$) leptoquarks $S_{2
\alpha}^{(\pm)}$ have the charge $\pm 2/3$. In general case
the scalar leptoquarks $S_{2 \alpha}^{(+)}$ and
$\starup{S_{2\alpha}^{(-)}}$
with electric charge 2/3 are mixed and can be
written as superpositions
\begin{eqnarray}
S_{2 \alpha}^{(+)}&=&\sum_{m=0}^3 c_m^{(+)}S_m, \; \; \; \; \; \;
 \starup{S_2^{(-)}}=\sum_{m=0}^3 c_m^{(-)}S_m \label{eq:mixS}
\end{eqnarray}
of three physical scalar leptoquarks $S_1$, $S_2$, $S_3$ with
electric charge 2/3 and a small admixture of the Goldstone mode
$S_0$. Here $c^{(\pm)}_{m}$, $m=0,1,2,3$ are the elements of the
unitary scalar leptoquark mixing matrix, $|c^{(\pm)}_{0}|^2=\frac
{1} {3}g_4^2 \eta_{3}^{2}/m_V^2 \ll 1$, $g_4$ is the $SU_V(4)$
gauge coupling constant, $\eta_3$ is the VEV of the
(15,2,1)-multiplet and $m_V$ is the vector leptoquark mass.

With account of the fermion mixing (\ref{eq:fmix}) the interaction of the
scalar leptoquark doublets (\ref{eq:SpSm}) with the fermions can be
described by the lagrangian
\begin{eqnarray}
{\cal L}_{SQl}  =
 (\,\, \linup{ {Q'}_{i a\alpha}^{L}} \,( h^{(+)}_l )_{ij} \, \, l^R_j \,)
 \, S_{a\alpha}^{(+)} \, +
(\,\, \linup{ {l'}_{ia}^{L}} \,( h^{(-)}_d )_{ij} \,\,  d^R_{j\alpha} \,)
 \, S_{a\alpha}^{(-)} \, +
\nonumber \\
(\,\, \linup{ {Q'}_{i a\alpha}^{L}} \,( h^{(-)}_{\nu} )_{ij} \,\, \nu^R_j \,)
 \,\tilde S_{a\alpha}^{(-)} \, +
(\,\, \linup{ {l'}_{ia}^{L}} \,( h^{(+)}_u )_{ij} \,\, u^R_{j\alpha} \,)
 \,\,\tilde S_{a\alpha}^{(+)} \, + \, h. \, c. , \label{eq:lagr}
\end{eqnarray}
where $ \tilde S_{a}^{(\pm)}=\varepsilon_{ab} \,\, \starup{S_{b}^{(\pm)}} $,
$\varepsilon_{12}=-\varepsilon_{21}=1$,
$\varepsilon_{11}=\varepsilon_{22}=0$ and

\begin{eqnarray}
 {Q'}_{i a\alpha}^{L} = \left ( \begin{array}{c}
u_{i \alpha}^{L}\\
(C_Q)_{ik} \, d_{k \alpha}^{L}
\end{array} \right ) , \,\,\,\,\,
 {l'}_{i a}^{L} = \left ( \begin{array}{c}
(C^{\dag}_l)_{ik} \, \nu_{k}^{L}\\
l_{i}^{L}
\end{array} \right )
\label{eq:Qlmix}
\end{eqnarray}
are the $SU_L(2)$ doublets of the left quarks and leptons,
$C_Q=(A_{Q_1}^{L})^{{\dag}}A_{Q_2}^L$ is the CKM-matrix,
$C_l=(A_{l_1}^{L})^{{\dag}}A_{l_2}^L$ is the analogous matrix in
the lepton sector and $(h^{(\pm)}_f)_{ij}$ are the coupling
constant matrixes,  the index $f$ denotes the right fermions
$f$ entering into (\ref{eq:lagr}).

In particular case of zero fermion mixing
($C_Q=C_l=I$, $(h^{(\pm)}_{f})_{ij}=h^{(\pm)}_{f_i}\delta_{ij}$)
the lagrangian (\ref{eq:lagr}) upon substitutions
 $S^{(+)}_{a}=\varepsilon_{ab} R_{2b}$,
 $S_{a}^{(-)}=(\tilde{R}_{2a})^*$,
$h^{(+)}_e=h_{2R}$,
 $h^{(-)}_d=(\tilde{h}_{2L})^*$, $h_u^{(+)}=-(h_{2L})^*$,
 $h_{\nu}^{(-)}=0$  reproduces the lagrangian of
ref. \cite{BRW},  $R_2$, $\tilde{R}_2$ and $h_{2L}$,
$h_{2R}$, $\tilde h_{2L}$ are the scalar leptoquarks and
the phenomenological coupling constants of ref. \cite{BRW}.

As a result of the Higgs splitting of the masses of quarks and leptons
 the general form
\cite{PovSm1} of Yukawa interactions of the scalar doublets with
the fermions in MQLS-model  \cite{AD1,AD2} gives for
the coupling constants of the lagragian (\ref{eq:lagr}) the
expressions
\begin{eqnarray}
(h_l^{(+)})_{ij} &=& -(C_Qh^R_2)_{ij} \, , \,\,\,\,
(h_d^{(-)})_{ij} = -(\krup{h^L_2})_{ij} \, , \\
\label{eq:g1}
(h_{\nu}^{(-)})_{ij} &=& -(h^R_1)_{ij} \, , \,\,\,\,
(h_u^{(+)})_{ij} = -(\krup{C_l} \krup{h^L_1})_{ij} \,
\label{eq:g2}
\end{eqnarray}
with
\begin{eqnarray}
(h^{L,R}_{a})_{ij}&=&\sqrt{{3 \over 2}}\frac{1}{\eta
\sin{\beta}} (M_{Q_a}K_a^{L,R}-K_a^{R,L} M_{l_a})_{ij} \, , \label{eq:g3}
\end{eqnarray}
where $(M_{f_a})_{ij}=m_{f_{ia}}\delta_{ij}$ are the diagonal mass
matrixes of quarks and leptons, $f_{ia}=Q_{ia}, \, l_{ia}$,
 $K^{L,R}_a=(A_{Q_a}^{L,R})^{{\dag}}A_{l_a}^{L,R}$ are the mixing
matrixes specific for the model with the four color quark-lepton
symmetry, $\beta$ is the angle of the mixing of the
(1,2,1)-doublet $\Phi^{(2)}$
with the 15th scalar doublet $\Phi^{(3)}_{15}$
of the (15, 2, 1)-- multiplet $\Phi^{(3)}$,
$\tan{\beta}=\eta_3/\eta_2$,
$\eta_2$ and $\eta_3$ are the VEVs of the $\Phi^{(2)}$ and
$\Phi^{(3)}$ multiplets, $\eta=\sqrt{\eta_2^2+\eta_3^2}$ is the SM VEV.

From the formulas (\ref{eq:lagr}) - (\ref{eq:g2})  we obtain
the responsible for the t-quark decays (\ref{eq:decayl}),
(\ref{eq:decaynu}) interactions of scalar leptoquarks
(\ref{eq:SpSm}), (\ref{eq:mixS}) with quarks and leptons  in the form

\begin{eqnarray}
L_{S^{(+)}_1Q_1l_2} &=& \bar Q_{i1\alpha } \Big [ h^{L}_{ij}P_{L}
+ h^{R}_{ij}P_{R} \Big ] l_{j2} S_{1\alpha}^{(+)} + {\rm h.c.},
\label{eq:lag1}\\
L_{S_{m}Q_{a}l_{a}} &=& \bar Q_{ia\alpha} \Big [
(h^{L}_{am})_{ij}P_{L}+(h^{R}_{am})_{ij}P_{R} \Big ] l_{ja}
S_{m\alpha} +{\rm h.c.} , \label{eq:lag2}
\end{eqnarray}
with
\begin{eqnarray}
(h^{L})_{ij} = (h^L_1 C_l)_{ij} \, , \,\,\,\,
(h^{R})_{ij} = -(C_Q h^R_2)_{ij} \, , \,\,\,\,
(h^{L,R}_{am})_{ij} = - (h^{L,R}_{a})_{ij} \, c^{L,R}_{am} \, ,
\label{eq:x5}
\end{eqnarray}
where the matrixes $h^{L,R,}_a$ are given by eq. (\ref{eq:g3}),
$c^{L,R}_{1m}=c^{(\pm)}_{m}$, $c^{L,R}_{2m}=c^{(\mp)}_{m}$,
 $c^{(\pm)}_{m}$ are the elements of the scalar leptoquark
mixing matrix in (\ref{eq:mixS}) and
$P_{L,R}=(1 \pm \gamma_5)/2$ are the left and right projection
operators.

 The interactions
(\ref{eq:lag1}), (\ref{eq:lag2}) induce in the tree approximation
the $t$-quark decays
\begin{eqnarray}
&&t \to u_i \; l_j^{+} \; l_k^{-} \, ,\label{eq:decayt1} \\
&&t \to u_i \; \tilde{\nu_j} \; \nu_k  \label{eq:decayt2}
\end{eqnarray}
with the production of the up quarks $u_i=(u,c), \, i=1,2$
according to the diagrams on Fig.1 
We have calculated the widths of the decays (\ref{eq:decayt1}),
(\ref{eq:decayt2}) with neglect of the final fermion masses
\begin{eqnarray}
  m_{u_i}, m_{l_j}, m_{\nu_{k}} \ll m_t, m_{S^{(+)}_1}, m_{S_m} \label{eq:con1}
\end{eqnarray}
and assuming that
\begin{eqnarray}
   m_{S^{(+)}_1}, m_{S_m} > m_t \, .  \label{eq:con2}
\end{eqnarray}

The widths of the decays (\ref{eq:decayt1})
are discribed by the diagram of Fig.1($a$) and
can be written in the form
\begin{eqnarray}
\Gamma(t \to u_i \, l^{+}_jl^{-}_k)=m_tH_{ij}H_{3k} \cdot
f_1(\mu_{S_1^{(+)}})/128(2\pi)^3 \; , \label{eq:width1}
\end{eqnarray}
where
\begin{eqnarray}
&&H_{i'j}=|h^{L}_{i'j}|^2+|h^{R}_{i'j}|^2 \; , \, i',j=1,2,3 \, , \label{eq:G} \\
&&f_1(\mu)  =  6\mu^2 - 5 - 2(\mu^2 - 1)(3\mu^2-1) \ln{ \frac
{\mu^2}{\mu^2-1}} \, \label{eq:f1}
\end{eqnarray}
and $ \mu_{S^{(+)}_1}= m_{S^{(+)}_1}/m_t \, .$ One can see from
(\ref{eq:x5}), (\ref{eq:g3}) that the widths (\ref{eq:width1}),
(\ref{eq:G}) are the largest ones for the heaviest final quark
$u_2=c$ and the corresponding largest coupling constants entering
the equation (\ref{eq:width1}), (\ref{eq:G}) can be approximately written as
\begin{eqnarray}
&&h^{L}_{i'j} \approx \sqrt{{3 \over 2}}\frac{m_{u_{i'}}}{\eta
\sin{\beta}}(K_1^LC_l)_{i'j} \label{eq:X1} \,\, , i'= 2, 3, \,\,
m_{u_{i'}}=m_c, m_t
\end{eqnarray}
(the coupling constants $h^{R}_{ij}$ are suppressed by the
non-diagonal matrix element of CKM matrix $C_Q$). As a result the
widths (\ref{eq:width1}), (\ref{eq:G}) for the decays
(\ref{eq:decayl}) are simplified and take the form
\begin{eqnarray}
\Gamma(t \to c \, l^{+}_j \, l^{-}_k)=m_t
\frac{\gamma_{tc}}{\sin^4{\beta}}k_{2j}k_{3k}f_1(\mu_{S_1^{(+)}})
\; , \label{eq:width2}
\end{eqnarray}
where
\begin{eqnarray}
\gamma_{tc}&=& \frac {9}{512(2\pi)^3} \cdot \frac {m_t^2m_c^2} {\eta^4} \, , \label{eq:gamma} \\
k_{i'j} & = & |(K_1^LC_l)_{i'j}|^2 \; , i',j=1,2,3 \label{eq:k} \,
.
\end{eqnarray}

The eq. (\ref{eq:width2}) predicts the $t \to c \, l^{+}_{j} \,
l_k^{-}$ decays with the production of the generation diagonal
charged lepton pairs $e^{+}e^{-}, \mu^{+}\mu^{-},
\tau^{+}\tau^{-}$ as well as of the non-diagonal ones such as
$e^{+}\mu^{-}, \mu^{+}e^{-}, e^{+}\tau^{-}, \dots$ in dependence
on the fermion mixing parameters $k_{2j}, \, k_{3k}$. Summarizing
the partial widths (\ref{eq:width2}) over the generation indexes
and using the unitarity of the matrixes $K_1^L, \, C_l$ we obtain
the total width of the charged lepton mode
\begin{eqnarray}
\Gamma(t \to c \, {l^{+}}' \, l^{-})= \sum_{j,k} \Gamma(t \to c \,
{l_j^{+}} \, l_k^{-})= m_t
\frac{\gamma_{tc}}{\sin^4{\beta}}f_1(\mu_{S_1^{(+)}}) \,
\label{eq:width3}
\end{eqnarray}
which is fermion mixing independent and includes all the decays
with the production of the every possible charged lepton pairs
both the diagonal and non-diagonal ones.  In particular case of
the zero fermion mixing ($K_1^LC_l=I$)
the total width (\ref{eq:width3}) is
saturated by the non-diagonal decay $t \to c \, \mu^{+} \,
\tau^{-}$ which is in this case the only allowed decay of type
(\ref{eq:decayl}).

The widths of the decays (\ref{eq:decayt2}) described by the
diagram of the Fig.~1($b$) are calculated with account of
(\ref{eq:lag2})  and can be written in the form
\begin{eqnarray}
\Gamma(t \to
u_i \, \tilde{\nu_j} \, \nu_k)=m_t\sum_{m,n}H^{mn}_{ij}H^{mn}_{3k} \cdot
f_2(\mu_{S_m},\mu_{S_n})/128(2\pi)^3 \; , \label{eq:width4}
\end{eqnarray}
where
\begin{eqnarray}
H^{mn}_{i'j}=(h_{1m}^{L})_{i'j}(h_{1n}^{L})^{*}_{i'j}+
(h_{1m}^{R})_{i'j}(h_{1n}^{R})^{*}_{i'j}, \, i',j=1,2,3 \, , \label{eq:G2} \\
f_2(\mu_1,\mu_2)=2(\mu_1^2+\mu_2^2)-3-2
\frac{\mu_1^2(\mu_1^2-1)^2}{\mu_1^2-\mu_2^2} \ln{ \frac
{\mu_1^2}{\mu_1^2-1}} \nonumber \\
-2\frac{\mu_2^2(\mu_2^2-1)^2}{\mu_2^2-\mu_1^2} \ln{ \frac
{\mu_2^2}{\mu_2^2-1}} \, \label{eq:f2}
\end{eqnarray}
and $\mu_{S_m}= m_{S_m}/m_t \, .$ Keeping in (\ref{eq:x5}), (\ref{eq:g3}) only
the largest terms proportional to $m_t$ and $m_c$
\begin{eqnarray}
&&(h^{L,R}_{1m})_{i'j} \approx -\sqrt{{3 \over 2}}\frac{m_{u_{i'}}}{\eta
\sin{\beta}}(K_1^{L,R})_{i'j} \, c_m^{(\pm)} \label{eq:Y1} \, , \,\, i'=2, 3,
\,\, m_{u_{i'}}=m_c, m_t
\end{eqnarray}
 we obtain from (\ref{eq:width4}), (\ref{eq:G2}) the
widths of the decays (\ref{eq:decaynu}) in the form
\begin{eqnarray}
\Gamma(t \to c \, \tilde{\nu}_j \, \nu_k)=m_t
\frac{\gamma_{tc}}{\sin^4{\beta}}\sum_{m,n=1}^{3}k^{mn}_{2j}k^{mn}_{3k}f_2(\mu_{S_m},
\mu_{S_n}) \, , \label{eq:width5}
\end{eqnarray}
where
\begin{eqnarray}
k^{mn}_{i'j}  =
|(K_1^L)_{i'j}|^2c^{*(+)}_mc^{(+)}_n+|(K_1^R)_{i'j}|^2c^{(-)}_mc^{*(-)}_n
\, , \, i',j=1,2,3 \label{eq:k2} \, ,
\end{eqnarray}
and $\gamma_{tc}$ is defined by the eq. (\ref{eq:gamma}).

Summarizing the partial widths (\ref{eq:width5}) over the
generation indexes and accounting the unitarity  of the matrixes
$K_1^{L,R}$ we obtain the total width of the neutrino mode
\begin{eqnarray}
\Gamma(t \to c\, \tilde{\nu}' \nu)= \sum_{j,k} \Gamma(t \to c \,
\tilde{\nu_j} \, \nu_k)= m_t
\frac{\gamma_{tc}}{\sin^4{\beta}}\sum_{m,n=1}^3
k^{mn}f_2(\mu_{S_m},\mu_{S_n}) \,  \label{eq:width6}
\end{eqnarray}
which is fermion mixing independent and contains the parameters
$k^{mn}=(c^{(+)}_m c^{*(+)}_n+c^{(-)}_m c^{*(-)}_n)^2$ depending
on the scalar leptoquark mixing (\ref{eq:mixS}). The expression
(\ref{eq:width6}) can be simplified in the particular case of the
scalar leptoquark mixing (\ref{eq:mixS}) with $c_3^{(\pm)}=0$
 when $S_2^{(+)}$ and
$\starup{S_{2}^{(-)}}$
are approximately the
superpositions of two physical scalar leptoquark \, $S_1$ and $S_2$
(the small admixture of the Goldstone mode can be neglected
because of the smallness of $c_0^{(\pm)}$). In this case $k^{mn}
\approx \delta_{mn}$ for $m,n=1,2$ and the width (\ref{eq:width6})
takes the form
\begin{eqnarray}
\Gamma(t \to c\, \tilde{\nu}' \nu)=m_t
\frac{\gamma_{tc}}{\sin^4{\beta}}[ f_1(\mu_{S_1})+f_1(\mu_{S_2})]
\, , \label{eq:width7}
\end{eqnarray}
here the relation $f_2(\mu,\mu)=f_1(\mu)$ has been taken also into
account.

The widths (\ref{eq:width3}), (\ref{eq:width7}) depend on the
masses $m_{S^{(+)}_1}$, $m_{S_1}$, $m_{S_2}$ of the scalar
leptoquarks and on the $\Phi^{(2)} - \Phi^{(3)}_{15} $ mixing angle
$\beta$.

The current data on the direct search for the leptoquarks set the
lower mass limits \cite{PDG04}
\begin{eqnarray}
 m_{LQ}>242 \, \mbox{GeV}, \, 202 \, \mbox{GeV}, 148 \,
\mbox{GeV} \label{dat:mass}
\end{eqnarray}
 for the scalar leptoquarks of the first,
of the second and the of third generation respectively. As
mentioned above the scalar leptoquarks $m_{S^{(+)}_1}$, $m_{S_1}$,
$m_{S_2}$ couple most intensively with t-quark. In the case
(\ref{eq:con2}) they will decay predominantly into $t \,
\tilde{l}_{ja}$ pairs and should be regarded as the third
generation ones. In this case the condition (\ref{eq:con2}) is
consistent with lower experimental limit 148 GeV in
(\ref{dat:mass}) and we use (\ref{eq:con2}) when choosing the
lower scale leptoquark masses in (\ref{eq:width3}),
(\ref{eq:width7}). With account of the total width of t-quark
$\Gamma_t^{tot} \approx \Gamma(t \to bW) \approx 1.56 \;
\mbox{GeV}$ we obtain from (\ref{eq:width3}), (\ref{eq:width7})
that
\begin{eqnarray}
&&Br (t \to c \, {l^{+}}' \, l^{-})=(5.7\cdot 10^{-8} - 0.6 \cdot
10^{-8} - 0.7\cdot 10^{-9})/ \sin^4{\beta} \, , \label{dat:br1} \\
&&Br (t \to c \, \tilde{\nu}' \, \nu)=(11.3\cdot 10^{-8} - 1.2 \cdot
10^{-8} - 1.4\cdot 10^{-9})/ \sin^4{\beta} \,  \label{dat:br2}
\end{eqnarray}
for $m_{S_1^{(+)}}, \,  m_{S_1}, \,  m_{S_2} = 180-250-400 \,
\mbox{GeV} \, .$

The mixing angle $\beta$ enters into the coupling constants
(\ref{eq:x5}),  (\ref{eq:g3}) and we restrict it by the smallness of
the pertubation theory parameter
\begin{eqnarray}
(h^{S,P})^2/4\pi < 1 \, , \label{eq:con3}
\end{eqnarray}
where $h^{S,P}=(h^{L} \pm h^{R})/2$ are the scalar and
pseudoscalar coupling constants and $h^{L,R}$ are the chiral
coupling constants (\ref{eq:x5}),  (\ref{eq:g3}). With account of
the $t$-quark chiral coupling constants in (\ref{eq:X1}), (\ref{eq:Y1})
as of the largest ones the condition (\ref{eq:con3}) gives that
$\sin{\beta}>0.12$.

The Fig.2 shows the branching ratio $Br(t \to c \, {l^{+}}' \,
l^{-})$ of the charged lepton mode as the function of the scalar
leptoquark mass $m_{S_1^{(+)}}=180-300~\mbox{GeV}$ for
a)~$\sin{\beta}=0.15$, b)~$\sin{\beta}=0.2$ and
c)~$\sin{\beta}=0.25$. The corresponding branching ratios $Br (t
\to c \, \tilde{\nu}'  \, \nu)$ of the neutrino mode for
$m_{S_1}=m_{S_2}=180-300~\mbox{GeV}$ are twice as large as those
of the charged lepton mode.

As is seen from the Fig.2 in all three cases a), b) and c) there
is the mass region with $Br(t \to c \, {l^{+}}' \, l^{-}) \sim 10^{-5}$.
For example for $\sin{\beta}=0.2$ from Fig.2-b and from
(\ref{dat:br1}), (\ref{dat:br2})  we obtain
\begin{eqnarray}
&&Br (t \to c \, {l^{+}}' \, l^{-})=(3.5 - 0.4 )\cdot 10^{-5} , \label{dat:br3} \\
&&Br (t \to c \, \tilde{\nu}' \, \nu)=(7.1 - 0.8) \cdot 10^{-5}
\label{dat:br4}
\end{eqnarray}
$\mbox{for} \,  m_{S_1^{(+)}}, \,  m_{S_1}, \,  m_{S_2} =
180-250~\mbox{GeV}$.
As is seen 
the branching ratios of the decays under consideration occur to be 
 of the same order as the sensitivity of LHC to the $t \to c X$ decays 
$Br(t \to c X) > 5 \cdot 10^{-5} $ \cite{F , B , GS, AB}.  

It would be of interest to estimate the sensitivity of LHC to the
decays~(\ref{eq:decayl}). It can be performed in an order-of-magnitude 
manner by using the studies \cite{Chik, Dodd} of LHC
sensitivity to the decays
\begin{eqnarray}
t \to c \, Z \to c \; l^+l^- , \label{eq:decaycZ}
\end{eqnarray}
where  $l^+l^-$ are both $e^+e^-$ and $\mu^+\mu^-$ pairs.



The decays (\ref{eq:decayl}) together with the decays  
\begin{eqnarray}
\tilde{t} \to \tilde{c} \; l^+_jl^-_k \label{eq:decayantit}
\end{eqnarray}
(the widths of the decays (\ref{eq:decayantit}) 
differ from (\ref{eq:width2}) by the transmutations 
$j \leftrightarrow k$ of the indices in the mixing parameters)
can manifest
themselves at LHC through the processes  
\begin{eqnarray}
p + p \to t \tilde{t} \to  \left \{ \begin{array}{c}
(c \,  l^+_jl_k^-)(W^-\tilde{b}) \to (c \, l^+_jl_k^-)(j j \tilde{b}) \\
(W^+b)(\tilde{c} \, l^+_jl_k^-) \to (j j \tilde{b})(\tilde{c} \, l^+_jl_k^-)
\end{array} \right \}  =
c' l^+_jl^-_kjj b', \label{eq:decayLHC}
\end{eqnarray}
where one of the quarks  ($t$ or $\tilde{t}$) decays according to
(\ref{eq:decayl}) or (\ref{eq:decayantit}) whereas the other one
($\tilde{t}$ or $t$) decays in the standard way followed by the
$W^{\mp} \to jj$ decays into two hadron jets. 
Although the case with leptonic decay modes $W \to l \nu_l$ 
is more sensitive \cite{Chik, Dodd} to the decays (\ref{eq:decaycZ})  
we use here for the analysys of the 
$t \to c \; l^+_jl^-_k$ decays  
the case with hadronic decays $W^{\mp} \to jj$ 
as the more simple one.  
In this case the final states are
$c'l^+_jl^-_kjjb'$ with $c'=c, \tilde{c}$ and $b'= \tilde{b}, b $ 
and the experimental signature includes, therefore, two charged leptons (in
general, of different generations) and four energetic jets.

For the generation diagonal signal process 
$$p + p \to t \tilde{t} \to
c'l^+l^-Wb' \to c'l^+l^-jjb' \to l^+l^- + 4\mbox{jets}$$ 
with $l^+l^-= e^+e^-, \, \mu^+ \mu^-$ 
the dominant backgrounds are the same as for
the process (\ref{eq:decaycZ}) 
\begin{eqnarray}
\nonumber && p + p \to Z + \mbox{jets} \to l^+l^-+ \mbox{jets}, \\  
\nonumber && p + p \to Z W \to l^+l^-+ \mbox{jets}, \\
\nonumber && p + p \to t \tilde{t} \to W^+bW^-\tilde{b} \to 
l^+\nu_{l}b l^-\tilde{\nu}_{l}\tilde{b} \to l^+l^-+ \mbox{jets}.
\label{eq:decayfon}
\end{eqnarray}
For the expected number of the signal events
$N^s_{l^+l^-}=N^s_{e^+e^-}+N^s_{\mu^+\mu^-}$ and for the background
ones we obtain the estimations
$$N^s_{l^+l^-}=1.1\cdot 10^7 \cdot Br(t \to c \, l^+l^-),$$
$$N^b_{l^+l^-}(Z+jets)=5.7\cdot 10^6, \,\,\,\,\, 
 N^b_{l^+l^-}(ZW)=1.3\cdot 10^4, \,\,\,\,\, 
 N^b_{l^+l^-}(t \tilde{t})=1.9\cdot 10^5 $$
for the expected LHC cross sections 
$\sigma_{t \tilde{t}} =833 pb$, $\sigma_{Zj} =8478 pb$, $\sigma_{ZW} =28 pb$
~\cite{Dodd} and for the integrated luminosity $L=10fb^{-1}$, 
here and below $Br(t \to c \, l^+l^-)=Br(t \to c \, e^+e^-) + Br(t \to c \, \mu^+\mu^-).$


To reduce the backgrounds one can use , in part, the cuts which
have been used for estimations of the sensitivity of LHC to the
decay (\ref{eq:decaycZ}). In the Table 1 we show the cuts which
have been used in ref. \cite{Dodd} in analysys of the events
$t\tilde{t} \to cZWb \to l^+l^- + 4 jets$ for the reconstraction 
of $t \to c Z$ decays. We also show the relative efficiencies
$\tilde{\varepsilon}^s$, $\tilde{\varepsilon}^b$ of each cut for
signal and backgrounds, which we have evaluated by using the results 
of ref. \cite{Dodd} where the detailed descriptions of the 
cuts can be also found.  

\begin{table}[h]
 \centerline{
\epsfxsize=1.0\textwidth \epsffile{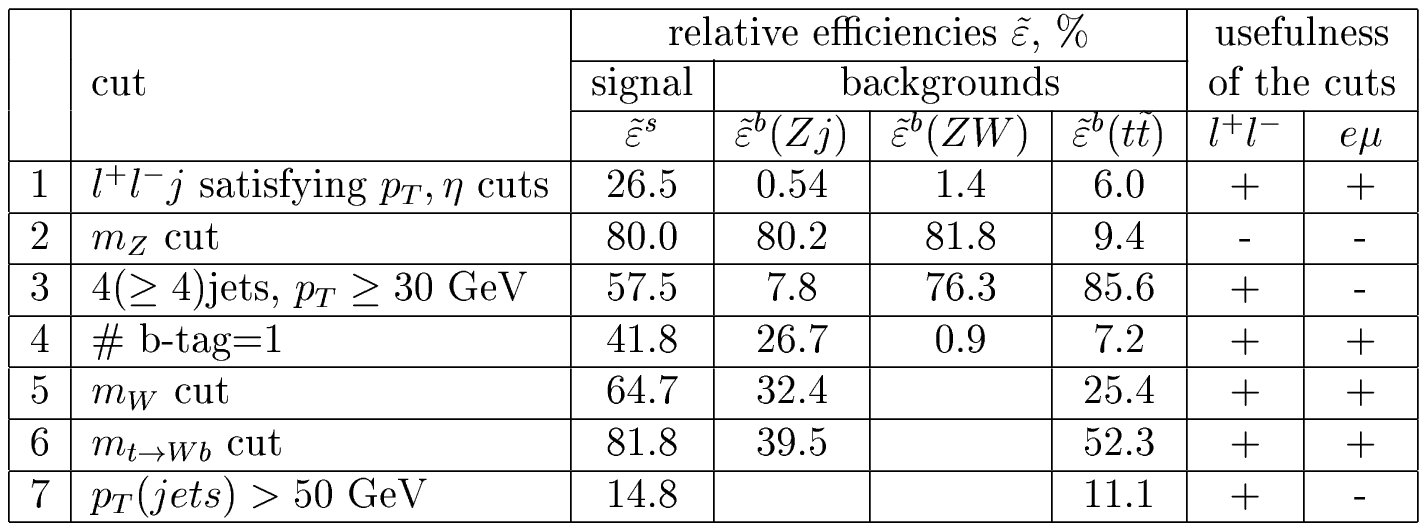}} \vspace*{1mm}
\label{tab:cuts}
\caption{Cuts used for reducing the backgrounds to the
  signal events  
  a)~$t \tilde{t} \to cZWb \to l^+l^-+4 jets$ \cite{Dodd}, \, \,  
  b)~$t \tilde{t} \to cl^+l^-Wb \to l^+l^-+4 jets$ and  
  c)~$t \tilde{t} \to ce\mu Wb \to e\mu+4 jets$. 
  $l^+l^-$ includes both $e^+e^-$ and $\mu^+\mu^-$ pairs,
  $e\mu$ includes both $e^-\mu^+$ and $e^+\mu^-$ ones. 
  The relative efficiencies $\tilde{\varepsilon}$ 
  are evaluated with using the results of ref.\cite{Dodd}.
  The last two column mark the cuts used in the cases b) and c). }
\end{table}

All the cuts are applicable to the analysis of the events 
$t \tilde{t} \to c'l^+l^-Wb' \to l^+l^- + 4jets$ except the $m_Z$ cut.
This cut requires the invariant mass of lepton pair to be near
$m_Z$ (within 4 GeV of $m_Z$) and it is not suitable for the decays 
$t\to c \, l^+l^-$. 
Assuming that after excluding the $m_Z$ cut the relative efficiencies 
of the remaining six cuts are unchanged 
and multiplying them we obtain the resulted efficiency
$\varepsilon^s_{l^+l^-}=5.0\cdot 10^{-3}$ for the signal events
and $\varepsilon^b_{l^+l^-} = \varepsilon^b_{l^+l^-}(t
\tilde{t})=5.4\cdot 10^{-5}$ for the background ones. As a result
of these cuts we can expect
$n^s_{l^+l^-}=\varepsilon^s_{l^+l^-}N^s_{l^+l^-}=5.6\cdot
10^{4}\cdot Br(t \to c \, l^+l^-)$ signal events and
$n^b_{l^+l^-}=n^b_{l^+l^-}(t \tilde{t})=\varepsilon^b_{l^+l^-}(t
\tilde{t})N^b_{l^+l^-}(t \tilde{t})\approx 10$ background ones.
Taking for the signal significance 
$s_{l^+l^-}=n^s_{l^+l^-}/ \sqrt{n^b_{l^+l^-}}$
the value $s_{l^+l^-}=5$ we obtain the expected sensitivity of LHC
to the $t \to c \, l^+ l^-$ decays
\begin{eqnarray}
Br(t \to c \, l^+ l^-)>3 \cdot 10^{-4} \label{dat:brtcll}
\end{eqnarray}
at $5 \sigma$ level for $L=10 fb^{-1}$.

For comparison, in the case of hadronic $W$ decay modes 
$W^{\mp} \to j j$    
the sensitivity of LHC to the decay 
$t \to c \, Z$   is   
$Br(t \to c \, Z)>1.7 \cdot 10^{-3}$  
\cite{Dodd} 
for $L=10 fb^{-1}$,     
which corresponds to the sensitivity  
$Br(t \to c \, Z \to c \; l^+l^- )>1.1 \cdot 10^{-4}$ 
to the decays (\ref{eq:decaycZ}).  
As is seen, the sensitivity (\ref{dat:brtcll}) 
is approximately by a factor three worse than that for the decay 
(\ref{eq:decaycZ}).

For the nondiagonal signal process
\begin{eqnarray}
\nonumber p + p \to t \tilde{t} \to c' e \mu Wb' \to c' e \mu j j b' \to e \mu +
4jets \label{eq:decay nondiag}
\end{eqnarray}
where $e\mu$ are both $e^-\mu^+$ and $e^+\mu^-$ pairs the dominant
background is given only by the process
\begin{eqnarray}
\nonumber p + p \to t \tilde{t} \to W^+bW^-\tilde{b} \to (e^+\nu_e
b)(\mu^-\nu_{\mu}\tilde{b})+(\mu^+\nu_{\mu}
b)(e^-\nu_{e}\tilde{b}) \to e \mu+jets.
\label{eq:decaynondiagback}
\end{eqnarray}
For numbers $N^s_{e \mu}=N^s_{e^- \mu^+}+N^s_{e^+ \mu_-}$ and
$N^b_{e\mu}=N^b_{e\mu}(t \tilde{t})$ of the signal and background
events we have the estimations
$$ N^s_{e\mu}= 1.1\cdot10^7\cdot Br(t \to c \, e \mu), \,\,\,\,\, 
 N^b_{e\mu}= 1.9\cdot10^5,$$
where $Br(t \to c \, e \mu)=Br(t\to c \, e^-\mu^+)+Br(t\to c \, e^+\mu^-).$

In the case of $e\mu$ pairs the $m_Z$ cut should be excluded, in
addition, the third cut and the seventh one (see the Table 1) 
can be also excluded
because these two cuts are suitable for reducing the $Zj$
background which is absent in the case of $e \mu$ events.
Assuming the relative efficiencies of the remaining four cuts
to be unchanged and multiplying  them we obtain the resulted efficiencies
$\varepsilon^s_{e\mu}=5.9\cdot10^{-2}$,
$\varepsilon^b_{e\mu}=\varepsilon^b_{e\mu}(t\tilde{t})=
5.7\cdot10^{-4}$ and the corresponding numbers
$n^s_{e\mu}=\varepsilon^s_{e\mu}N^s_{e\mu}=6.6\cdot 10^{5}\cdot 
Br(t\to c e \mu)$, $n^b_{e\mu}=\varepsilon^b_{e\mu}N^b_{e\mu} \approx
108$ of the signal and background events. As a result for the 
signal significance $s_{e\mu}=5$ we can expect the LHC sensitivity
to the $t \to c e \mu$ decay 
\begin{eqnarray}
Br(t \to c e \mu)>8\cdot10^{-5} \label{dat:brlimit}
\end{eqnarray}
at $5\sigma$ level for $L=10fb^{-1}$.

The approximate estimations (\ref{dat:brtcll}),
(\ref{dat:brlimit}) relate to the case of hadronic $W$ decay modes 
$W^{\mp} \to j j$ and correspond to an integrated luminosity
$L=10fb^{-1}$. These estimations can be improved approximately 
by order of magnitude 
\begin{eqnarray}
Br(t \to c \, l^+l^-)\gtrsim 10^{-5}, \; \; \; Br(t \to c e \mu)\gtrsim 10^{-6}
\label{dat:brsim}
\end{eqnarray}
for $L=100fb^{-1}$ and, possibly, the further improvement can be achieved 
by accounting the leptonic $W$ decay modes $W \to l \nu_l$.     

As is seen the branching ratios (\ref{dat:br3}), (\ref{dat:br4}) 
are of the same order as the expected sensitivities (\ref{dat:brsim}) 
of LHC to the decays $t \to c \, l^+l^-$ 
and $t \to c \, e \mu.$    
The search for these decays at LHC (especialy for the decays of 
$t \to c \, e \mu$ type having the more clean signature) 
may be of interest. The detection of such decays
would be the clear sign of the new physics,
possibly, induced by the four color symmetry between quarks and
leptons. 

It should be noted that for the masses 180 - 300 GeV 
the leptoquarks can be directly produced at LHC and 
by the additional studies of their decay modes, branching ratios etc. 
it will be possible to to clean up the origin of the observed leptoquarks. 
In this situation the detection of the 
$t \to c \, l^{+}_j \, l^{-}_k$ 
decays could be an additional argument in favour of 
the scalar leptoquarks discussed in this paper.   
It is worth noting that the non-diagonal decays of 
$t \to c \, l^{+}_j \, l^{-}_k$
type are also predicted in the general two Higgs doublet model 
but with the essentially less branching ratios, for example with 
$Br(t \to c \;  \tau^{-} \mu^{+})\sim 10^{-8}$ 
as the largest one \cite{UT}.

In conclusion we resume the results of the work. The rare $t$-quark
decays $t \to c \, l^{+}_j \, l^{-}_k$, $t \to c \, \tilde{\nu}_j
\, \nu_k$ induced by the scalar leptoquark doublets are
investigated in the minimal four color symmetry model with the
Higgs mechanism of the quark-lepton mass splitting. The partial
widths $\Gamma(t \to c \, l^{+}_j \, l^{-}_k)$ and $\Gamma(t \to c
\, \tilde{\nu}_j \, \nu_k)$ of these decays are calculated in tree
approximation and the total width of the charged lepton mode
$\Gamma(t \to c \, {l^{+}}' \, l^{-})=\sum_{j,k}\Gamma( t \to c \,
l^{+}_j \, l^{-}_k)$ and the neutrino one $\Gamma(t \to c \,
\tilde{\nu}' \, \nu)=\sum_{j,k}\Gamma( t \to c \, \tilde{\nu}_j \
\nu_k)$ are found in the fermion mixing independent form. The
corresponding branching ratios $Br (t \to c \, {l^{+}}' \,
l^{-})$, $Br (t \to c \, \tilde{\nu}' \, \nu)$ are shown to be of
order of $10^{-5}$ for the scalar leptoquark masses $ \,
m_{S_1^{(+)}}, \, m_{S_1}, \, m_{S_2} = 180-250 $ GeV and for
$\sin{\beta} \approx 0.2$, $\beta$ is the $\Phi^{(2)} - \Phi^{(3)}_{15}$
mixing angle of the model. These estimations are close
to the possible sensitivity of LHC to these decays and the search
for the decays $t \to c \, l^{+}_j \, l^{-}_k$, $t \to c \,
\tilde{\nu}_j \, \nu_k$ at LHC may be of interest.

\vspace{3mm} {\bf Acknowledgments}

The work was partially supported by the Russian Foundation for
Basic Research under grant 04-02-16517-a.

\newpage

{\Large\bf Figure captions}

\bigskip

\begin{quotation}

\noindent
Fig. 1. The diagrams of the rare $t$-quark decays
    a)~$t \to c \, l^{+}_j \, l_{k}^{-}$
    and b)~$t \to c \, \tilde{\nu}_j \, \nu_{k}$
        via scalar leptoquarks $S^{(+)}_1$ and $S_m$,
    $m=1,2,3$ of the MQLS-model. \\

\noindent
Fig. 2. The branching ratio $Br(t \to c \, {l^{+}}' \, l^{-})=
    \sum_{j,k} Br(t \to c \,{l_j^{+}} \, l_k^{-})$
        of the charged lepton mode as a function of the
    scalar leptoquark mass
        $ m_{S^{(+)}_1} $ for a)~$\sin{\beta}=0.15$,
    b)~$\sin{\beta}=0.20$, c)~$\sin{\beta}=0.25$. \\

\end{quotation}

\newpage
\begin{figure}[htb]
\vspace*{0.5cm}
 \centerline{
\epsfxsize=1.0\textwidth \epsffile{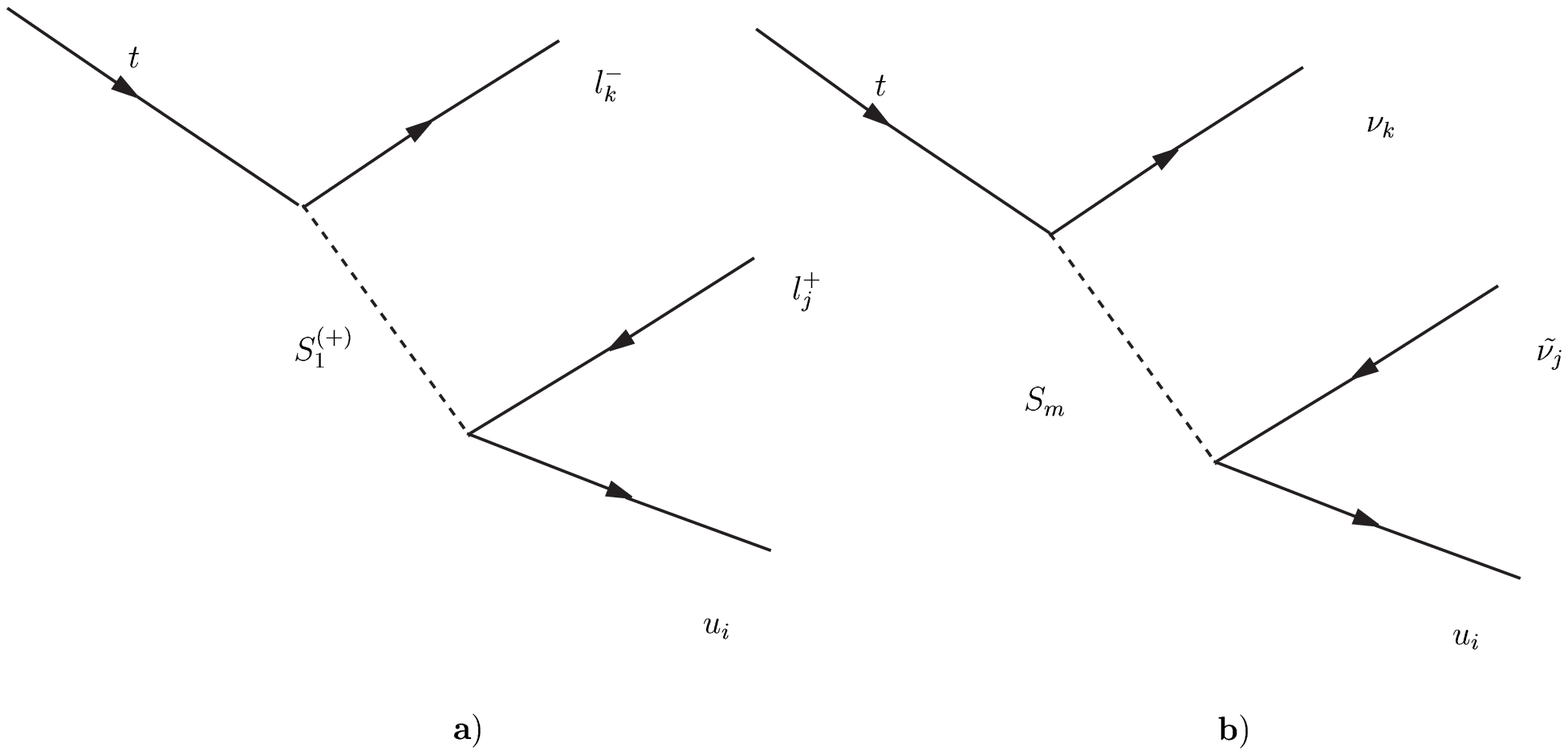}} \vspace*{1mm}
\label{fig:diag}
\end{figure}

\vfill \centerline{P.Yu.~Popov, A.D.~Smirnov, Modern Physics Letters A}

\centerline{Fig. 1}

\newpage
\begin{figure}[htb]
\vspace*{0.5cm}
 \centerline{
\epsfxsize=.80\textwidth \epsffile{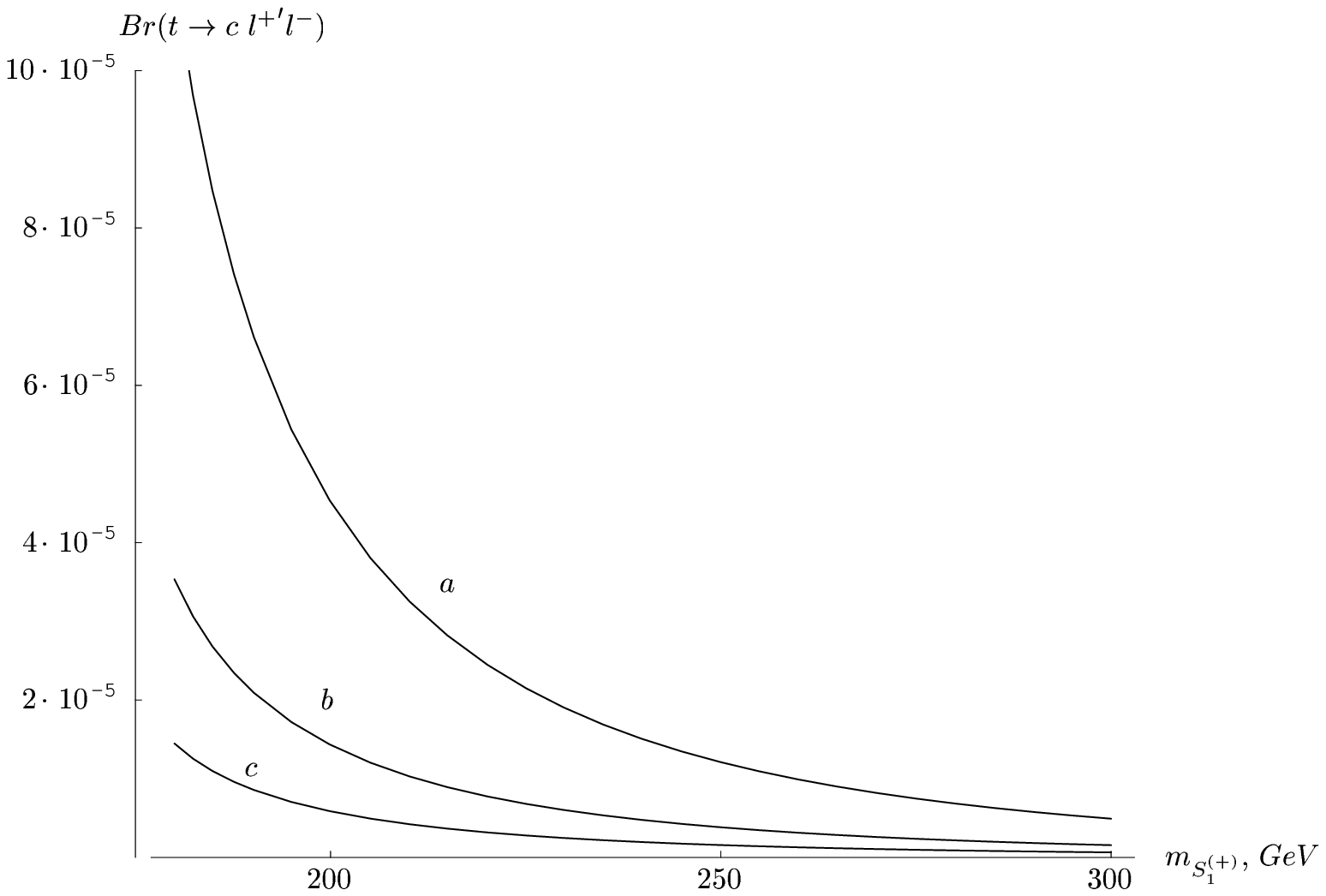}} \vspace*{1mm}
\label{fig:grafic}
\end{figure}

\vfill \centerline{P.Yu.~Popov, A.D.~Smirnov, Modern Physics Letters A}

\centerline{Fig. 2}

\end{document}